\documentclass[twocolumn,preprintnumbers,amsmath,amssymb,aps,prb,superscriptaddress]{revtex4-2}

\pdfoutput=1
\usepackage[pdftex]{graphicx}
\usepackage[english]{babel}
\usepackage{soul}
\usepackage{cleveref}
\usepackage{bbold}
\usepackage{mathtools}
\usepackage{multirow}
\usepackage{colortbl}
\usepackage{tikz}
\definecolor{kugray5}{RGB}{224,224,224}
\usepackage[latin1]{inputenc}
\usepackage{diagbox}
\usepackage{mciteplus}
\usepackage{appendix}

\usepackage{mathrsfs}
\usepackage{relsize}

\usepackage{dcolumn}
\usepackage{bm}

\usepackage{color}
\usepackage{amstext}
\usepackage{braket}
\usepackage{mathdots}
\usepackage{tikz}
\usepackage{physics}
\usepackage{enumitem}

\usetikzlibrary{matrix,decorations.pathreplacing}

\begin{document}


\title{Generalized Lieb's theorem for noninteracting non-Hermitian $n$-partite tight-binding lattices}

\author{A. M. Marques}
\email{anselmomagalhaes@ua.pt}
\affiliation{Department of Physics $\&$ i3N, University of Aveiro, 3810-193 Aveiro, Portugal}

\author{R. G. Dias}
\affiliation{Department of Physics $\&$ i3N, University of Aveiro, 3810-193 Aveiro, Portugal}



\begin{abstract}
Hermitian bipartite models are characterized by the presence of chiral symmetry and by Lieb's theorem, which derives the number of zero-energy flat bands of the model from the imbalance of sites between its two sublattices.
Here, we introduce a class of non-Hermitian models with an arbitrary number of sublattices connected in a unidirectional and cyclical way and show that the number of zero-energy flat bands of these models can be found from a generalized version of Lieb's theorem, in what regards its application to noninteracting tight-binding models, involving the imbalance between each sublattice and the sublattice of lowest dimension.
Furthermore, these models are also shown to obey a generalized chiral symmetry, of the type found in the context of certain clock or parafermionic systems. 
The main results are illustrated with a simple toy model, and possible realizations in different platforms of the models introduced here are discussed.
\end{abstract}


\maketitle
\section{Introduction}
\label{sec:intro}
Lieb's theorem \cite{Lieb1989}, initially formulated to demonstrate that the ground state magnetization at half filling of repulsive Hubbard bipartite lattices is directly proportional to the sublattice (SL) imbalance \cite{Shen1994,Gouveia2015,Gouveia2016b,Tindall2021}, is now understood in a more broad sense.
Concretely, it states that the number of zero-energy flat bands (FBs) of a crystalline and bipartite tight-binding (TB) model, of which Lieb-type lattices are a prime example \cite{Morales2016,Zhang2017,Madail2019,Mao2020,Ni2020}, is given by the SL imbalance \cite{Ezawa2020,Marques2021,Marques2021b}. In real-space, the global sublattice imbalance of any bipartite system (including non-crystalline ones) indicates the lower bound on the number of zero-energy states present there \cite{Kikutake2013}.

Within the context of non-Hermitian systems, several studies have already addressed the formation, persistence or destruction of FBs in these models through different approaches \cite{Ge2018,Maimaiti2021}, most commonly with the introduction of parity-time ($\mathcal{PT}$) symmetric perturbations \cite{Ge2015,Lazarides2019,Jin2019}, including in one-dimensional (1D) \cite{Molina2015,Xia2021,Li2022a} and two-dimensional (2D) \cite{Zhang2019} Lieb-type lattices.
The compact localized states associated with these FBs have already been experimentally detected in a $\mathcal{PT}$-symmetric photonic trimer chain \cite{Biesenthal2019} with balanced gains and losses.
Tuning the parameters of these systems to fall on exceptional points has been shown to drive the formation of FBs \cite{Ramezani2017,Leykam2017}.
Here, we introduce a certain class of non-Hermitian models with  $n\geq 2$ SLs, which we call $n$-partite systems, and show that the number of zero-energy FBs in these models is given by a generalized version of Lieb's theorem, as it is understood in the specific context of the noninteracting TB models studied here.
This constitutes a novel mechanism for the formation of FBs in non-Hermitian systems, which is not dependent on any specific symmetries, like $\mathcal{PT}$ symmetry, even though the models studied here have a built-in generalized chiral symmetry by default.

Bipartite models are also characterized by the presence of chiral symmetry, which pairs eigenvalues with symmetric energies.
Some extensions of the usual chiral symmetry have already been considered, whether for $q$-deformed Hamiltonians \cite{Kawarabayashi2011,Kawarabayashi2016,Kawarabayashi2021}, in 1D models with finite energy edge states topologically protected by a chiral-like symmetry \cite{Marques2019}, in models with different adiabatically connected chiral symmetry representations at different liming cases \cite{Dias2022}, or in 1D superlattices \cite{Marques2020}, not necessarily bipartite, with point-chiral symmetry \cite{Anastasiadis2022} whose energy spectrum is symmetric about a finite momentum value.
Models belonging to the class introduced here, on the other hand, are shown to obey the same generalized chiral symmetry as the one found in the generalized quantum Ising chains known as Baxter's clock models \cite{Baxter1989,Fendley2014}.
A simple 1D toy model is introduced for the purpose of illustrating both the generalized Lieb's theorem and the generalized chiral symmetry.

The rest of the paper is organized as follows.
In Sec.~\ref{sec:npartite}, we define $n$-partite models and introduce their general Hamiltonian. 
Then, we discuss the symmetries of these models, with a particular emphasis on the generalized chiral symmetry. 
We end this section by formally deriving a generalized version of Lieb's theorem, which counts the total number of zero-energy FBs  in these non-interacting $n$-partite TB lattices.
In Sec.~\ref{sec:toy}, we introduce a toy model that exemplifies the main results found in the previous section.
We also analyze the energy spectrum of this toy model for open boundaries, showing that the FB states survive the emergence of the skin effect, while the dispersive states do not.
Finally, we present our conclusions in Sec.~\ref{sec:conclusions}.

\section{$n$-partite models} 
\label{sec:npartite}
\begin{figure}[ht]
	\begin{centering}
		\includegraphics[width=0.48 \textwidth,height=3.3cm]{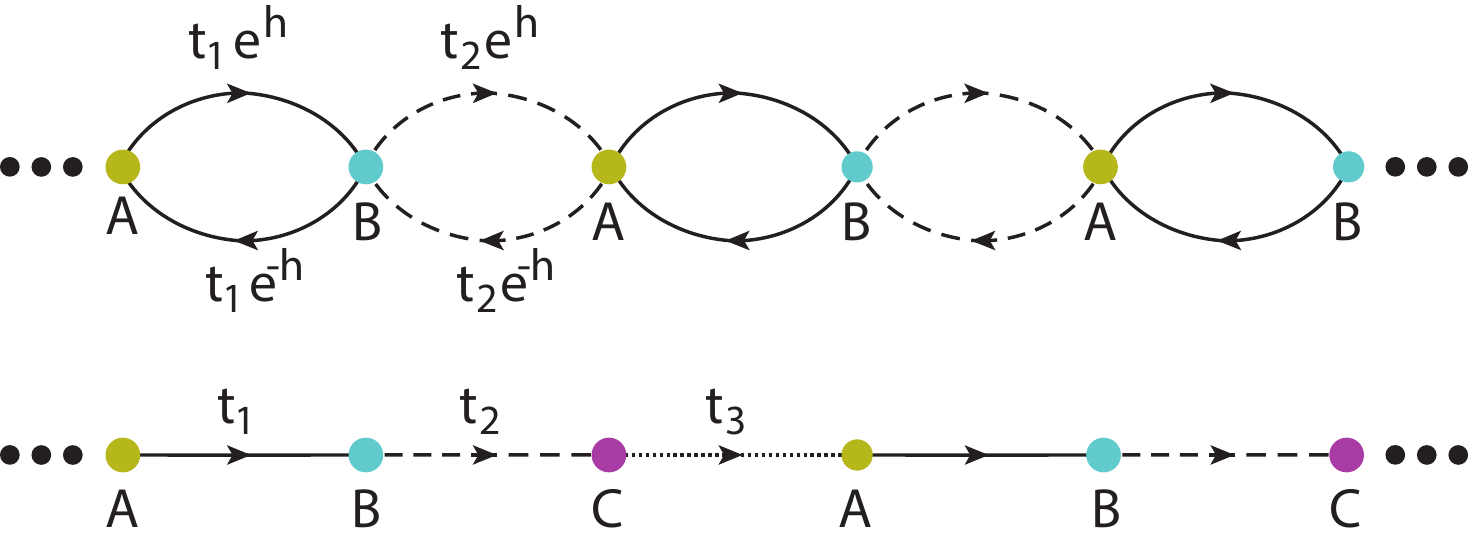} 
		\par\end{centering}
	\caption{Top: non-Hermitian bipartite model with staggered hoppings.  Bottom: tripartite trimer chain. The arrows indicate the direction of the hopping parameters and $h\in\mathbb{R}$, corresponding to an imaginary gauge field.}
	\label{fig:6}
\end{figure}
We define an $n$-partite model as a system composed of $n$ SLs, where each SL, in turn, is defined as a group of sites that can only connect between themselves through integer multiples of $n$-hopping processes.
This implies that only bipartite (2-partite) lattices can be Hermitian (the $h=0$ case in Fig.~\ref{fig:6} top), while for $n>2$ the model is necessarily non-Hermitian and built with \textit{unidirectional} couplings (see Fig.~\ref{fig:6} bottom), which further imposes the absence of $\mathcal{PT}$ symmetry (since an inversion operation also inverts the direction of the couplings).
Let us consider the general form of an $n$-partite Hamiltonian with $n$ SLs of arbitrary sizes, each coupled to its neighbor in a directed fashion,
\begin{equation}
	H(\mathbf{k})=
	\begin{pmatrix}
		&h_1&&&
		\\
		&&h_2&&
		\\
		&&&\ddots&
		\\
		&&&&h_{n-1}
		\\
		h_n&&&&
	\end{pmatrix},
\label{eq:hamiltnroot}
\end{equation}  
where the entries not shown are zeros, the momentum vector reads as $\mathbf{k}=(k_1,k_2,\dots,k_D)$, with $D$ the dimensionality of the system, and $h_j=h_j(\mathbf{k})$, with $j=1,2,\dots,n$, is a rectangular matrix of size $d_j\times d_{j+1}$, with $j=n+1\to j=1$ from the periodic boundary conditions. This Hamiltonian describes a periodic model composed of unidirectional hopping terms from sites in SL$_j$ to sites in SL$_{j-1}$.
Note that, according to our definition, the Hamiltonian of \textit{all} possible $n$-partite models can be written either in the form of (\ref{eq:hamiltnroot}) or as its conjugate transpose version, $H(\mathbf{k})\to H^\dagger(\mathbf{k})$, which corresponds to a global inversion of all coupling directions.
Upon raising the Hamiltonian in (\ref{eq:hamiltnroot}) to the $n^\text{th}$ power, one arrives at a diagonal matrix of the form
\begin{eqnarray}
	H^n(\mathbf{k})&=&\text{diag}(H_1,H_2,\dots,H_n), 
	\label{eq:hamiltnrootnpower}
	\\
	H_j&=&h_{j}h_{j+1}\dots h_{n-1+j}.
	\label{eq:hamiltnrootblockl}
\end{eqnarray}
Each diagonal block $H_j$ is a $d_j\times d_j$ square matrix, and the set $\{H_j\}$ represents all cyclic permutations of the ordered product of all the original $h_j$ matrices. 
We note that there is a recent study in driven systems \cite{Zhou2022} where the authors, by considering Floquet operators with formal properties similar to those of (\ref{eq:hamiltnroot})-(\ref{eq:hamiltnrootblockl}), were able to construct high-root \cite{Dias2021,Marques2021,Marques2021b,Bomantara2021,Deng2022} Floquet topological insulators of any order.

We assume for convenience that the SLs are ordered in a way that obeys $d_1\leq d_{j\neq 1}$, such that $H_1$ in (\ref{eq:hamiltnrootnpower}) is the smallest block (or in the set of smallest blocks) of dimension $d_1\times d_1$.
It can then be shown that the energy spectrum of $H_1$ is shared by all other $H_{j\neq 1}$, such that it is $n$-fold degenerate in $H^n(\mathbf{k})$. The Schr\"{o}dinger equation for the $H_1$ block is written as
\begin{equation}
	H_1\ket{u^1_s(\mathbf{k})}=E_{1,s}(\mathbf{k})\ket{u^1_s(\mathbf{k})},\  \ s=1,2,\dots,d_1,
	\label{eq:schrodh1}
\end{equation}
where $\ket{u^1_s(\mathbf{k})}$ is the eigenstate with momentum $\mathbf{k}$ of band $s$, and only has weight on the $d_1$ components of the first sublattice.
Applying $h_n$ on both sides of (\ref{eq:schrodh1}) and using the identity $h_{j}H_{j+1}=H_{j}h_{j}$, derived from (\ref{eq:hamiltnrootblockl}), leads to
\begin{equation}
	H_n\big(h_n\ket{u^1_s(\mathbf{k})}\big)=E_{1,s}(\mathbf{k})\big(h_n\ket{u^1_s(\mathbf{k})}\big),
\end{equation}
which, after defining the $d_n$-dimensional (non-normalized) eigenvector $\ket{u^n_s(\mathbf{k})}:=h_n\ket{u^1_s(\mathbf{k})}$, becomes
\begin{equation}
	H_n\ket{u^n_s(\mathbf{k})}=E_{1,s}(\mathbf{k})\ket{u^n_s(\mathbf{k})},\  \ s=1,2,\dots,d_1.
	\label{eq:schrodhn}
\end{equation}
Since $d_n\geq d_1$, (\ref{eq:schrodhn}) only accounts for the $d_1$ energy bands that are proven to be degenerate with the equivalent ones coming from the diagonalization of $H_1$. There are, however, extra $d_n-d_1$ bands coming from $H_n$ which do not belong to the shared spectrum.
From a sequential application of $h_{n-1,},h_{n-2},\dots,h_2$ to both sides of (\ref{eq:schrodhn}), one can generalize this proof to show that
\begin{equation}
	H_j\ket{u^j_s(\mathbf{k})}=E_{1,s}(\mathbf{k})\ket{u^j_s(\mathbf{k})},\  \ s=1,2,\dots,d_1,
	\label{eq:schrodhl}
\end{equation}
where the $d_j$-dimensional (non-normalized) eigenvectors are defined as $\ket{u^j_s(\mathbf{k})}:=h_j\ket{u^{j+1}_s(\mathbf{k})}$, which only have weight on SL$_j$.

\subsection{Generalized chiral symmetry}
\label{subsec:genchiral}

In the absence of gauge fields, the spinless fermionic Hamiltonian in (\ref{eq:hamiltnroot}) obeys both complex conjugation symmetry (corresponding to the time-reversal symmetry, $\mathscr{T}$, for Hermitian systems) and a generalized version of particle-hole ($\mathscr{P}_n$) symmetry, defined respectively as
\begin{eqnarray}
	\mathscr{T}&:&\ \ \ \ TH(\mathbf{k})T^{-1}_n=H(\mathbf{-k}),\ \ \ \ \ \ \ \ \ \ T=K,
	\label{eq:trsym}
	\\
	\mathscr{P}_n&:&\ \ \ P_nH(\mathbf{k})P^{-1}_n=\omega_n^{-1}H(\mathbf{-k}),\ \ \ P_n=\Gamma_nK ,
	\label{eq:genphsym}
	\\
	\Gamma_n&=&\text{diag}(\mathbb{1}_{d_1},\omega_n\mathbb{1}_{d_2},\dots,\omega_n^{n-2}\mathbb{1}_{d_{n-1}},\omega_n^{n-1}\mathbb{1}_{d_n}),
	\label{eq:genchiraloperator}
\end{eqnarray} 
where $K=K^{-1}$ represents the complex conjugation operation, with $KK^{-1}=1$, $\mathbb{1}_{d_j}$ is the identity matrix of dimension $d_j$, $\Gamma_n\Gamma_n^{-1}=\Gamma_n^n=\mathbb{1}_{d_H}$, with $d_H=\sum_{j=1}^{n}d_j$ the dimension of $H(\mathbf{k})$, $\omega_n=e^{i\frac{2\pi}{n}}$ and $\omega_n^n=1$.
From the combination of these two symmetries one can also define a generalized chiral ($\mathscr{C}_n$) symmetry, whose generalized chiral operator is written as $C_n=P_nT=\Gamma_n$,
\begin{equation}
	\mathscr{C}_n:\ \ \ \Gamma_nH(\mathbf{k})\Gamma^{-1}_n=\omega_n^{-1}H(\mathbf{k}),
	\label{eq:genchiralsym}
\end{equation} 
which constitutes another branch on the already long list of non-Hermitian symmetries, as systematically studied in \cite{Kawabata2019}.
For a bipartite system, $n=2$, (\ref{eq:genchiralsym}) reduces to the usual chiral symmetry \cite{Yin2018,Yao2018,Yokomizo2019} defined as $\Gamma_2H(\mathbf{k})\Gamma_2^{-1}=-H(\mathbf{k})$.
Therefore $\Gamma_n$ is the operator defining the chiral symmetry of an $n$-partite system, defined by the presence of $n$ sublattices.
It should be stressed that, since  $\Gamma_n$ is unitary, the system retains its $	\mathscr{C}_n$ symmetry even in the presence of gauge fields, that is, even when both $	\mathscr{T}$ and $	\mathscr{P}_n$ symmetries are broken (as is the case for the usual chiral symmetry in bipartite lattices crossed by finite magnetic fluxes \cite{Li2022}).

The presence of $\mathscr{C}_n$ symmetry imposes a constraint on the complex energy spectrum of $H(\mathbf{k})$.
Let us consider a \textit{finite} energy eigenstate of the system,
\begin{equation}
	H(\mathbf{k})\ket{\psi_0(\mathbf{k})}=E\ket{\psi_0(\mathbf{k})},
\end{equation}
then, by iteratively applying $\Gamma_n$ on both sides and using (\ref{eq:genchiralsym}) at each iteration, one arrives at
\begin{equation}
	H(\mathbf{k})\ket{\psi_l(\mathbf{k})}=\omega_n^lE\ket{\psi_l(\mathbf{k})},
	\label{eq:schrodgenchiral}
\end{equation}
where $\ket{\psi_l(\mathbf{k})}:=\Gamma_n^l\ket{\psi_0(\mathbf{k})}$, $l=1,2,\dots,n-1$, and $\braket{\psi_i(\mathbf{k})}{\psi_j(\mathbf{k})}=\delta_{ij}$.
This tells us that if $E$ is a finite eigenvalue of an eigenstate of the system, then all its rotated versions, given by the $n-1$ sequential $\phi_n=\frac{2\pi}{n}$ rotations in the energetic Argand plane, are also eigenvalues of orthogonal eigenstates, \textit{i.e.}, the finite eigenvalues come in sequences of the form $\{E,\omega_nE,\omega_n^2E,\dots,\omega_n^{n-1}E\}$, and the values in each sequence sum to zero.
This can also be understood by directly developing (\ref{eq:genchiralsym}) as
\begin{equation}
	\Gamma_n^lH(\mathbf{k})\Gamma^{-l}_n=\omega_n^{-l}H(\mathbf{k}),\ \ \ \ l=1,2,\dots,n-1,
\end{equation}
which, in particular, implies that $n$-partite lattices also obey the generalized chiral symmetry of all the divisors of $n$.
For example, for a $6$-partite lattice one not only has $\mathscr{C}_6$ symmetry, but also $\mathscr{C}_3$ and $\mathscr{C}_2$ symmetries, whose operators are given by $\Gamma_3=\Gamma_6^2$ and $\Gamma_2=\Gamma_6^3$, respectively.
As a corollary, all even-partite lattices possess the usual chiral symmetry $\mathscr{C}_2$.
In analogy with the colored states that can be present in certain $XXZ$ Heisenberg models \cite{Lee2020,Chertkov2021}, we can similarly identify the action of $\Gamma_n$ on $H(\mathbf{k})$ as an ordered transformation between different \textit{chiral colors} of the same Hamiltonian, defined from (\ref{eq:genchiralsym}) and for $n=3$, e.g., as
\begin{eqnarray}
	H_{\tikz\draw[red,fill=red] (0,0) circle (.3ex);}&=&H(\mathbf{k}),
	\label{eq:hred}
	\\
	H_{\tikz\draw[blue,fill=blue] (0,0) circle (.3ex);}&=&\Gamma_3H_{\tikz\draw[red,fill=red] (0,0) circle (.3ex);}\Gamma_3^{-1}=\omega_3^{-1}H_{\tikz\draw[red,fill=red] (0,0) circle (.3ex);},
	\label{eq:hblue}
	\\
	H_{\tikz\draw[green,fill=green] (0,0) circle (.3ex);}&=&\Gamma_3H_{\tikz\draw[blue,fill=blue] (0,0) circle (.3ex);}\Gamma_3^{-1}=\omega_3^{-2}H_{\tikz\draw[red,fill=red] (0,0) circle (.3ex);},
	\label{eq:hgreen}
\end{eqnarray}
with $\tikz\draw[red,fill=red] (0,0) circle (.3ex);\overset{\Gamma_3}{\to}\tikz\draw[blue,fill=blue] (0,0) circle (.3ex);\overset{\Gamma_3}{\to}\tikz\draw[green,fill=green] (0,0) circle (.3ex);\overset{\Gamma_3}{\to}\tikz\draw[red,fill=red] (0,0) circle (.3ex);$, from where it can be seen that
\begin{equation}
		H_{\tikz\draw[red,fill=red] (0,0) circle (.3ex);}+H_{\tikz\draw[blue,fill=blue] (0,0) circle (.3ex);}+	H_{\tikz\draw[green,fill=green] (0,0) circle (.3ex);}=0,
		\label{eq:sumcolorsh3}
\end{equation}
that is, and in more general terms, the $n$ chiral colors of a given $\mathscr{C}_n$-symmetric Hamiltonian sum to zero (or, alternatively, the $n$ colors sum to white).

The operator of the generalized chiral symmetry was first introduced in the context of the tripartite Hermitian breathing kagome model \cite{Ni2019}, and shown to pin the higher-order corner modes at zero energy. However, it has been recently demonstrated that the $\mathscr{C}_3$ symmetry of the model fails to protect the corner modes against certain perturbations that preserve it \cite{Miert2020}.
On the basis of the energetic constraints imposed by $\mathscr{C}_n$ symmetry on the specific non-Hermitian Hamiltonians of the form of (\ref{eq:hamiltnroot}), and encapsulated in (\ref{eq:schrodgenchiral}), we argue that the Hermitian models obeying $\mathscr{C}_3$ \cite{Ni2019,Kempkes2019,Ezawa2022,Herrera2022,Anastasiadis2022} or  $\mathscr{C}_4$ \cite{Li2020b,Li2021} symmetry studied so far fail to reveal the relevant consequences of the generalized chiral symmetry detailed here (a more expanded discussion can be found in Appendix~\ref{app:genchiralsym}). 
That is because $\mathscr{C}_n$-symmetric Hermitian models require $n$ applications of the generalized chiral symmetry in order to recover the original Hamiltonian, whereas in our case the Hamiltonian is recovered, up to a global phase factor [see (\ref{eq:genchiralsym})], after each application of the symmetry transformation, \textit{i.e.}, acting with $\Gamma_n$ on the Hamiltonian changes its chiral color.
The class of models introduced here, namely non-Hermitian $n$-partite models with unidirectional hopping terms between adjacent SLs defined in a cyclic fashion, should be regarded as the \textit{first example of a fully $\mathscr{C}_n$-symmetric class of TB models}.

It should be noted, however, that the same kind of generalized chiral symmetry has already been addressed in a different context, namely, that of generalized quantum Ising chains known as Baxter's clock model \cite{Baxter1989,Baxter1989b}, where the ``spin'' or clock internal degree of freedom at each site can take any value $\omega_n^j$, with $j=0,1,\dots,n-1$ (this model can be reframed in a parafermionic language, as shown, e.g., in \cite{Fendley2014,Alicea2016}).
A brief introduction to Baxter's clock model is provided in Appendix~\ref{app:baxter}, along with the analogies that can be drawn between this model and the one introduced in this paper.

It is convenient to introduce the \textit{phase commutator} between matrices $A$ and $B$, which we define as
\begin{equation}
	[A,B]_\theta :=AB-e^{-i\theta}BA,\ \ \ \ \ \theta\in[0,2\pi),
	\label{eq:phasecomm}
\end{equation}
reducing to the commutation relation for $[A,B]_0=[A,B]$, and to the anti-commutation relation for $\theta=\pi$, $[A,B]_\pi=\{A,B\}$.
The compact expression for $\mathscr{C}_n$ in (\ref{eq:genchiraloperator}) can be restated, through (\ref{eq:phasecomm}), as a phase commutator of the form
\begin{equation}
	[\Gamma_n,H(\mathbf{k})]_{\phi_n}=0,
	\label{eq:phasecommgenchiral}
\end{equation}
where, in particular, one recovers the known anticommutation relation for a bipartite model as $\{\Gamma_2,H(\mathbf{k})\}=0$, while also trivially recovering the commutation relation $[\Gamma_1,H(\mathbf{k}]=0$, since $\Gamma_1=\mathbb{1}$.
In the context of Baxter's clock model analyzed in Appendix~\ref{app:baxter}, the phase commutator in (\ref{eq:phasecommgenchiral}) can be viewed as the analog of the ``$\omega$ commutator'' \cite{Baxter1989,Fendley2014} of the generalized Clifford algebra \cite{Zhou2022} involving the local operators with which the Hamiltonian of this model is constructed.

To conclude the discussion of the symmetries of $n$-partite models, let us consider a Hermitian Hamiltonian that can be written as $H^\prime(\mathbf{k})=H(\mathbf{k})+H^\dagger(\mathbf{k})$, where $H(\mathbf{k})$ is given in (\ref{eq:hamiltnroot}).
We further assume that $H^\prime(\mathbf{k})$ has inversion or parity symmetry, written as $RH^\prime(\mathbf{k})R^{-1}=RH(\mathbf{k})R^{-1}+RH^\dagger(\mathbf{k})R^{-1}=H^\prime(-\mathbf{k})$, where $R$ is the inversion operator.
It is then straightforward to see that the following identity holds,
\begin{equation}
	RH(\mathbf{k})R^{-1}=H^\dagger(-\mathbf{k}),
\end{equation}
which can be seen as a modified inversion symmetry for the $n$-partite model.
More concretely, if $H^\prime(\mathbf{k})$, constructed from $H(\mathbf{k})$, has inversion symmetry, then the latter can be said to enjoy inversion symmetry also, \textit{up to a global inversion of the hopping directions}.
\begin{figure}[ht]
	\begin{centering}
		\includegraphics[width=0.34 \textwidth,height=4cm]{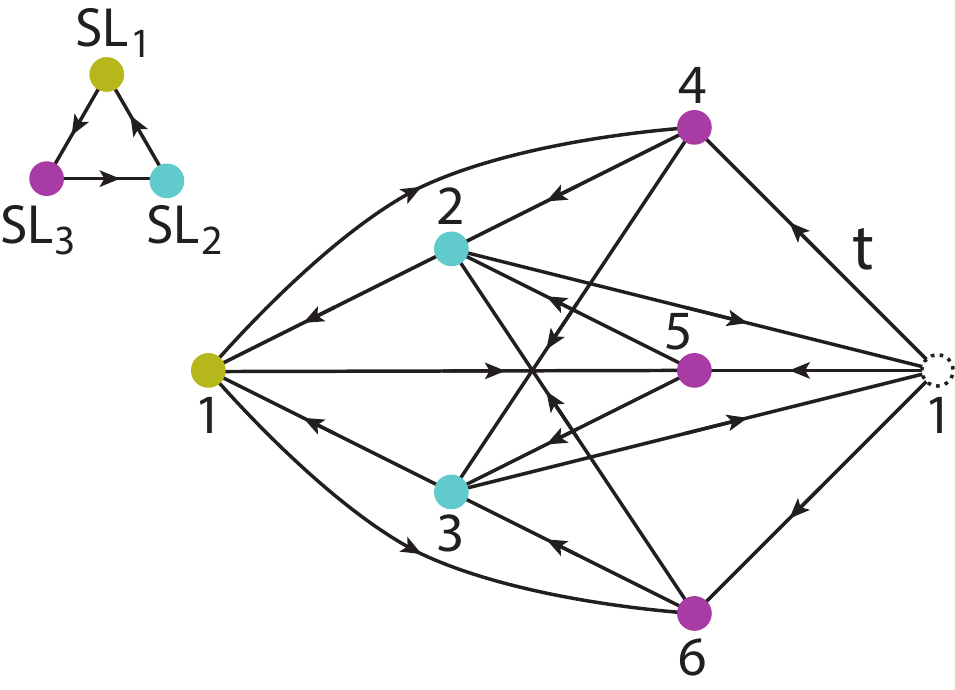} 
		\par\end{centering}
	\caption{Unit cell of the toy model for $n=3$. The hopping terms are unidirectional and follow the direction of the arrows. Open site 1 at the right belongs to the adjacent unit cell. The flow between sublattices is depicted at the top left.}
	\label{fig:1}
\end{figure}

\subsection{Generalized Lieb's theorem}
\label{subsec:genlieb}
The combination of the results above leads to another important result. 

(i) On the one hand, from the discussion leading to (\ref{eq:schrodhl}), we found that the finite energy spectrum of the smallest diagonal block of $H^n(\mathbf{k})$, which we set as $H_1$, is $n$-fold degenerate. 
On the other hand, this translates in the original model $H(\mathbf{k})$, through (\ref{eq:schrodgenchiral}), as an $n$-fold degeneracy of all \textit{absolute} finite energy values, which also leads to an $n$-fold degeneracy of $E^n$.
As a result, whenever $d_1\leq d_{j>1}$, the extra bands of $H_j$, in relation to $H_1$, must be zero-energy FBs, otherwise their finite energies would have to be
$n$-fold degenerate, that is, shared \textit{also} by $H_1$, which is not the case.

(ii) Furthermore, if $H_1$ is itself bipartite, \textit{i.e.}, if there are $\#_{\text{FB}}^{H_1}$ zero-energy FBs in the spectrum of $H_1$ coming from sublattice imbalance \textit{within} SL$_1$, then the same number of extra FBs appears in the other $n-1$ diagonal blocks $H_{j>1}$, meaning that the degenerate block spectra given by (\ref{eq:schrodhl}) actually remains valid for zero-energy FBs, that is, when $E_{1,s}(\mathbf{k})=0$.
The same reasoning of (i) can be applied here to prove the negative is impossible.
Let us suppose that we construct $\#_{\text{FB}}^{H_1}$  dispersive bands in each of the $H_{j>1}$ blocks, with global $(n-1)$-fold degeneracy for band $E_\alpha(\mathbf{k})>0$, with $\alpha=1,2,\dots,\#_{\text{FB}}^{H_1}$ the band index.
Then, due to the $\mathscr{C}_n$-symmetry of the original Hamiltonian, the finite energies appear in groups of $n$ elements of the form $\{\sqrt[n]{E_\alpha(\mathbf{k})},\omega_n\sqrt[n]{E_\alpha(\mathbf{k})},\dots,\omega_n^{n-1}\sqrt[n]{E_\alpha(\mathbf{k})}\}$.
However, $E_\alpha(\mathbf{k})$ was assumed to be $(n-1)$-fold degenerate, since it is absent from $H_1$, and therefore cannot originate the $n$ elements for each $\mathbf{k}$ mentioned above for the original Hamitonian and, as a consequence, the extra $(n-1)\#_{\text{FB}}^{H_1}$ bands of the spectrum are also zero-energy FBs, that is, degenerate with the $\#_{\text{FB}}^{H_1}$ FBs present in $H_1$ \footnote{When $\#_{\text{FB}}^{H_1}=2$, e.g., one could believe it would be possible to construct, from the $2n-2$ remaining bands from the other $H_{j>1}$ blocks, an $n$-fold degenerate finite energy band plus an $(n-2)$-fold degenerate zero-energy FB. However, the eigenstates of a finite energy band cannot vanish at any SL, which includes SL$_1$ (it can be easily checked that, if there are nodes at all sites of one SL, then, through the TB equations, these nodes propagate sequentially to all sublattices). As such, if the eigenstates of this finite energy band have finite weight on some sites of SL$_1$, then it would have to be present in $H_1$ also, which is assumed not to be the case. Therefore all $2n-2$ extra bands are also zero-energy FBs.}.

Since the number of zero-energy FBs is the same for $H^n(\mathbf{k})$ and $H(\mathbf{k})$, the results of this subsection can be summarized in the following formula that generalizes Lieb's theorem for a $\mathscr{C}_n$-symmetric $n$-partite system:
\begin{equation}
	\#_{\text{FB}}=\sum\limits_{j=2}^n(d_j-d_1)+n\#_{\text{FB}}^{H_1},
	\label{eq:genlieb}
\end{equation}
that is, the total number of zero-energy FBs is given by the sum of imbalances between each SL$_{j> 1}$ and the smallest sublattice SL$_1$ [the first term on the right, coming from point (i) above], plus $n$ times the number of zero-energy FBs already present in the smallest $H_1$ block [the second term on the right, coming from point (ii) above].
It should be noted that the second term should be included already for non-Hermitian systems with $n=2$, as we illustrate in Appendix~\ref{app:4rootcl} with an example, showing that Lieb's theorem can be generalized also for the (bipartite) lattices for which it was formulated.
As a corolary, we can also infer that if $H_1$ has a real energy spectrum, all $H_{j\neq 1}$ have real spectra, given that their extra bands must be zero-energy FBs, such that they are pseudo-Hermitian Hamiltonians \cite{Mosta2002} obeying $H_{j>1}^\dagger=\eta H_{j>1}\eta^{-1}$, with $\eta$ a positive definite unitary matrix \cite{Zhang2021} that reduces to the identity for Hermitian Hamiltonians.

One should be reminded, at this point, that $H(\mathbf{k})$ in (\ref{eq:hamiltnroot}) is non-Hermitian and therefore can be defective, that is, the number of linearly independent eigenstates (LIEs) of $H(\mathbf{k})$ can be lower than its dimensionality $d_H$, if $H(\mathbf{k})$ falls into  exceptional points or lines of the parameter space \cite{Bergholtz2021}.
Regarding the eigenstates, (\ref{eq:genlieb}) should be interpreted as giving the maximum possible number of LIEs within the set of zero-energy FBs of $H(\mathbf{k})$. 
However, defective models can have less LIEs in this set than $\#_{\text{FB}}$, down to a minimum given by
\begin{equation}
	\#_{\text{LIEs}}^{\text{min}}=\sum\limits_{j=2}^n \text{Max}(d_j-d_{j-1},0),
\end{equation}
which we derive in Appendix~\ref{app:defective}, where an explicit example of a defective system is also provided.

\section{Toy model}
\label{sec:toy}
\begin{figure*}[t]
	\begin{centering}
		\includegraphics[width=0.975 \textwidth,height=4.1cm]{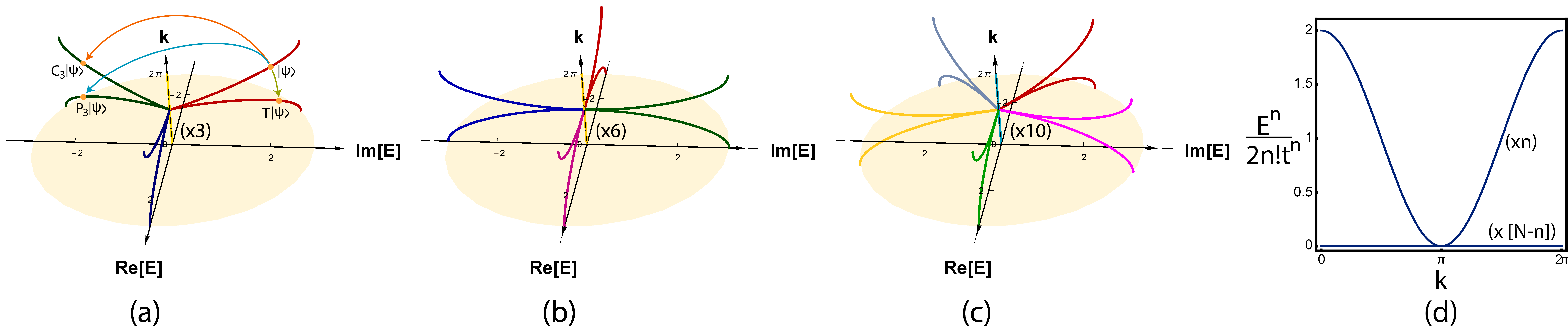} 
		\par\end{centering}
	\caption{Complex energy spectrum as a function of the momentum obtained from diagonalizing the Hamiltonian defined in (\ref{eq:toyh1})-(\ref{eq:toyhn}) for (a) $n=3$, (b) $n=4$, and (c) $n=5$. 
		The different symmetric partners of state $\ket{\psi}$ are indicated in (a).
		(d) Normalized energy spectrum as a function of the momentum of the model in (a)-(c) raised to the $n^\text{th}$ power, which is purely real. In all plots, ($\times j$) indicates the $j$-fold degeneracy of the respective band, $N=\sum_{i=1}^n i$ is the total number of bands, and only the zero-energy FB is degenerate in (a)-(c).}
	\label{fig:2}
\end{figure*}
In order to illustrate the results above, we introduce the simple 1D $n$-partite model ($\mathbf{k}\to k$), with a bulk Hamiltonian of the form of (\ref{eq:hamiltnroot}), whose entries are explicitly given by
\begin{eqnarray}
	h_1&=&t(1+e^{-ik})J_{1\times 2},
	\label{eq:toyh1}
	\\
	h_j&=&tJ_{j\times j+1},\ \ \ \ \ \ \ \ \ \ \ \  j=2,3,\dots,n-1,
	\\
	h_n&=&t(1+e^{ik})J_{n\times 1},
	\label{eq:toyhn}
\end{eqnarray}
where $J_{i\times j}$ is a matrix of ones of size $i\times j$, the lattice spacing was set to $a\equiv1$ here and everywhere below, and $t$ is the magnitude of the unidirectional hopping terms, set as the energy unit henceforth.
The unit cell of this model for $n=3$ is depicted in Fig.~\ref{fig:1}.
When raised to the $n^\text{th}$ power, this Hamiltonian has the form of (\ref{eq:hamiltnrootnpower}), with the diagonal blocks reading as
\begin{equation}
	H_j=2\frac{n!}{j}t^n(1+\cos k)J_{j\times j},
	\label{eq:toyhtonblocks}
\end{equation}
with $j=1,2,\dots,n$, such that $H_j$ is a matrix of size $j\times j$, that is, $d_j=j$ is the number of sites in SL$_j$. In particular, the smallest block is already diagonal and has the form $H_1=2n!t^n(1+\cos k)$, which models a simple uniform and Hermitian linear chain with hopping strength $n!t^n$ and an overall $2n!t^n$ energy shift.
The energy band characterizing the spectrum of $H_1$ is $n$-fold degenerate in $H^n(k)$ through (\ref{eq:schrodhl}), since it is common to all $H_j$ blocks.

The energy spectrum of the model defined through (\ref{eq:toyh1})-(\ref{eq:toyhn}) is shown for different $n$ in Figs.~\ref{fig:2}(a)-\ref{fig:2}(c).
The presence of the respective $\mathscr{C}_n$ symmetry is apparent in all three cases, as the energy spectra are manifestly invariant under $\phi_n$ rotations about the $k$ axis.
We illustrate the $\mathscr{T}$, $\mathscr{P}_3$, and $\mathscr{C}_3$ symmetric partners of an arbitrary state $\ket{\psi}$ in Fig.~\ref{fig:2}(a).
At the same time, the degeneracy of the zero-energy FBs agrees with the generalized Lieb's theorem expressed in (\ref{eq:genlieb}).
In Fig.~\ref{fig:2}(d), we show the normalized energy spectrum of $H^n(k)$, whose diagonal blocks are given by (\ref{eq:toyhtonblocks}).
Notice that this spectrum is purely real since the smallest block $H_1$ has a real spectrum, and that the degeneracy of zero-energy FB reconfirms the generalized Lieb's theorem, which can also be checked against the independent diagonalization of all $H_{j>1}$ blocks and counting the total number of FBs each of them generates.

\subsection{Open boundary conditions}
	\label{subsec:obc}
\begin{figure}[ht]
	\begin{centering}
		\includegraphics[width=0.48 \textwidth,height=8.cm]{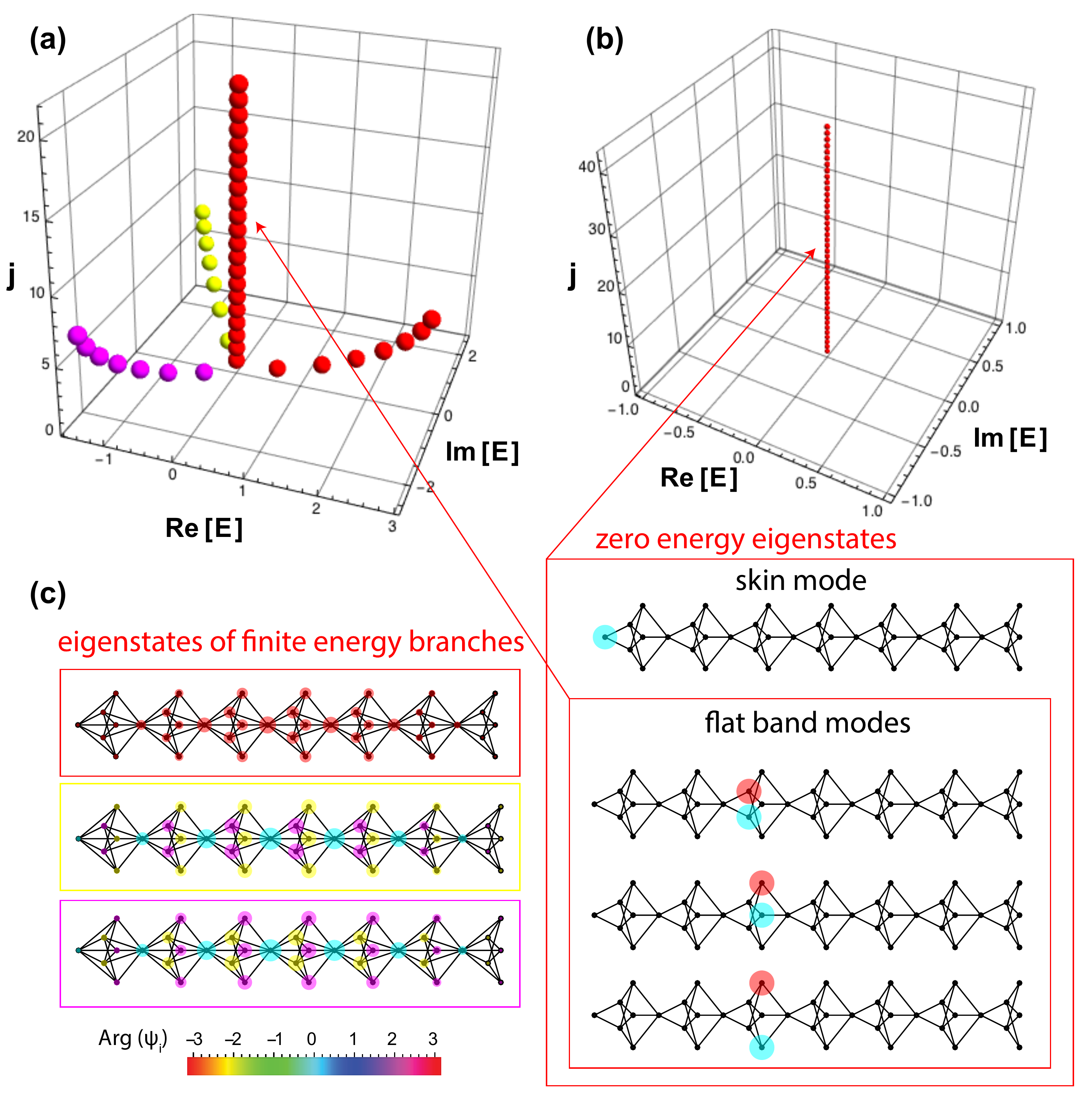} 
		\par\end{centering}
	\caption{Complex energy spectrum, in units of $t$, of the toy model of Fig.~\ref{fig:1} under OBC and with $N=7$ unit cells as a function of the state index $j$ for (a) the complete model, and (b) the model without loops, that is, without all rightwards directed hopping terms. The finite energy eigenstates of the branches in (a) become the skin modes of (b), while the number of zero-energy states of the FBs remains the same for both cases. (c) Examples of eigenstate profiles of $\mathscr{C}_3$ symmetric partners at the left, and of the skin mode and three different zero-energy FB states at the right, where the radius of the circle represents the amplitude of the wavefunction at the respective site and the color represents its phase, coded by the color bar below.}
	\label{fig:toymodelobc}
\end{figure}

In this section, we briefly discuss the effects of considering open boundary conditions (OBC) for the toy model of Fig.~\ref{fig:1}, both with and without closed loops.
In Fig.~\ref{fig:toymodelobc}(a), we plot the complex energy spectrum of this 3-partite toy model under OBC and for $N=7$ unit cells.
Three finite energy branches of seven states each can be observed, with an example of a $\mathscr{C}_3$ symmetric triplet given at the left of Fig.~\ref{fig:toymodelobc}(c), together with 21 zero-energy FB states, in agreement with the bulk spectrum of Fig.~\ref{fig:2}(a).
Interestingly, the skin effect is absent from this system, even though it is composed of non-Hermitian unidirectional couplings.
The reason for this is that \textit{loops} are present in the configuration of the hopping terms, which prevents the eigenstates from converging to a given edge.
Even though unidirectionality is assumed for the couplings, the toy model is built in such a way that there is no dominant hopping direction, with a global balance between leftwards and rightwards oriented hopping terms.

In Fig.~\ref{fig:toymodelobc}(b), we plot the same complex energy spectrum as in Fig.~\ref{fig:toymodelobc}(a), only removing from the open chain all the rightwards directed hopping terms (see Fig.~\ref{fig:1}), such that there are no loops formed by the hopping terms.
Immediately we see that only the finite energy states of the three branches are affected, namely by having all of them collapsing into the zero-energy skin mode with weight at the left edge site only, as depicted at the top right of Fig.~\ref{fig:toymodelobc}(c).
At the same time, it can be seen that the number of zero-energy FB states is unaltered.
This is to be expected, since each FB mode can be written as a compact state that only has weight on the SL from which it is derived, as exemplified for the three FB state depicted at the right in Fig.~\ref{fig:toymodelobc}(c). 
As we discuss in Appendix~\ref{app:defective}, a Gram-Schmidt orthogonalization can be applied to the chain such that the FB modes become isolated sites in the rotated basis.
Therefore the case of Fig.~\ref{fig:toymodelobc}(b) highlights the fact that the FB states are insensitive to the boundary conditions of the system, while the eigenstates corresponding to the dispersive bulk bands for PBC may or may not coalesce into skin modes under OBC, depending on the presence or absence of loops in the model.

\section{Conclusions}
\label{sec:conclusions}
We introduced a class of non-Hermitian TB models characterized by the presence of $n\geq 2$ SLs coupled in a cyclic fashion through unidirectional couplings. 
When the Hamiltonian of these models is raised to the $n^\text{th}$ power it becomes block diagonal, with the energy spectrum of the smallest block, corresponding to the $H_1$ block in (\ref{eq:hamiltnrootnpower}) by construction, being a common feature of all blocks.
The excess energy bands of the blocks of higher dimensionality, in relation to the smallest one, were proved to be zero-energy FBs.
The same total number of these zero-energy FBs is also present at the original Hamiltonian, which enabled us to generalize Lieb's theorem \cite{Lieb1989}, originally only applicable to Hermitian bipartite systems, to account for the total number of these bands in the non-Hermitian $n$-partite models we considered.

At the same time, we showed how these models obey a generalized chiral symmetry $\mathscr{C}_n$ of the type introduced in \cite{Ni2019}.
On the basis of the different action of this symmetry on our TB models and on those appearing in recent literature \cite{Ni2019,Kempkes2019,Li2020b,Ezawa2022,Li2021,Anastasiadis2022}, we argued that only the former can be properly characterized as $\mathscr{C}_n$-symmetric models, expanding their class beyond the generalized spin systems for which this symmetry was originally proposed \cite{Baxter1989,Fendley2014}.
A toy model was introduced to illustrate the appearance of zero-energy FBs whose cardinality is in agreement with the generalized Lieb's theorem, and to manifest the $n$-fold rotation symmetry of the complex energy spectrum (see Fig.~\ref{fig:2}), which is a direct consequence of the $\mathscr{C}_n$ symmetry.

Since the $n^\text{th}$ root of the energy spectrum of the $H_1$ block was shown to be $n$-fold degenerate, in absolute value, at the level of the original Hamiltonian, the development of techniques to control the exact form of $H_1$ can open up interesting perspectives.
To name only one, if $H_1$ is dressed with topological features by appropriately designing the original model $H(\mathbf{k})$, then the latter will inherit its topological characterization directly from the former.
In other words, one can use this method to construct $\mathscr{C}_n$-symmetric $n$-root topological insulators, which are currently limited to $2^n$-root systems \cite{Dias2021,Marques2021,Marques2021b}, therefore extending to nondriven systems the recent results obtained for Floquet insulators \cite{Zhou2022}. 
These results are being finalized and will be the subject of a forthcoming article \cite{Marques2022}.

Concerning the experimental realization of the non-Hermitian $n$-partite models studied here, the main challenge relates to the implementation of unidirectional hopping terms.  
In this regard, electrical circuits appear to be in a prominent position to realize these systems \cite{Hofmann2019,Zhang2022,Zeng2022}, since unidirectional capacitance couplings can be designed with the use of impedance converters with current inversion, which have already been shown to be experimentally accessible \cite{Liu2021b,Zou2021}.
Unidirectional couplings can  also be very well approximated in systems where a strong imaginary gauge field can be induced, since these translate as highly asymmetrical nonreciprocal couplings.
This can be achieved  in quantum systems such as (i) photonic lattices, either with ring resonators coupled by mediating auxiliary rings with balanced gains and losses \cite{Longhi2015,Longhi2015b,Longhi2018} or, as has been experimentally realized recently, with light walks in photonic fibers \cite{Weidemann2020,Weidemann2022}, (ii) ultracold atoms in optical lattices, where similar protocols based on exploiting the effects of transitions to an auxiliary lattice to generate highly asymmetric hopping terms have been proposed \cite{Gong2018,Liu2019,He2021}.
At the same time, imaginary gauge fields have also been implemented in classical setups, namely, by including auxiliary acoustic cavities with air dissipative materials in acoustic lattices \cite{Zhang2021b}, or even in robotic metamaterials \cite{Branden2019}, where the lattice can be mapped into a system of masses coupled by springs with effective nonreciprocal spring constants.

\section*{Acknowledgments}
\label{sec:acknowledments}

This work was developed within the scope of the Portuguese Institute for Nanostructures, Nanomodelling and Nanofabrication (i3N) projects No.~UIDB/50025/2020 and No.~UIDP/50025/2020 and funded by FCT - Portuguese Foundation for Science and Technology through the Project No. PTDC/FIS-MAC/29291/2017. AMM acknowledges financial support from the FCT through the work Contract No.~CDL-CTTRI-147-ARH/2018 and from i3N through the work Contract No. CDL-CTTRI-46-SGRH/22.
The authors would like to thank David Viedma for useful discussions and suggested bibliographic material.

\appendix

\section{Comments on the generalized chiral symmetry}
\label{app:genchiralsym}

We start by considering a Hamiltonian of the form
\begin{equation}
	H_0(\mathbf{k})=
	\begin{pmatrix}
		0&h_{12}&\omega_3 h_{13}
		\\
		h_{12}^*&0&\omega_3^2 h_{23}
		\\
		\omega_3 h_{13}^*&\omega_3^2 h_{23}^*&0
	\end{pmatrix},
\end{equation}
where $\omega_3=e^{i\frac{2\pi}{3}}$ and all $h_{ij}=h_{ij}(\mathbf{k})$ are scalars.
This model was introduced in \cite{Ezawa2022} to model $\mathbb{Z}_3$ clock parafermions in a breathing kagome lattice.
This Hamiltonian can be decomposed as (the momentum dependence is omitted henceforth)
\begin{eqnarray}
	H_0&=&H_\circlearrowright+H_\circlearrowleft,
	\\
	H_\circlearrowright&=&
	\begin{pmatrix}
		0&h_{12}&0
		\\
		0&0&\omega_3^2 h_{23}
		\\
		\omega_3 h_{13}^*&0&0
	\end{pmatrix},
	\\
	H_\circlearrowleft&=&
	\begin{pmatrix}
		0&0&\omega_3 h_{13}
		\\
		h_{12}^*&0&0
		\\
		0&\omega_3^2 h_{23}^*&0
	\end{pmatrix}.
\end{eqnarray}
The generalized chiral symmetry $\mathscr{C}_3$ defined in (\ref{eq:genchiralsym}) reads here as
\begin{equation}
	\Gamma_3H_0\Gamma^{-1}_3=\omega_3^{-1}H_\circlearrowright+ \omega_3 H_\circlearrowleft :=H_1.
\end{equation}
The action of $\Gamma_3$ is therefore to produce two counterpropagating $\phi_3$ rotations on the Hamiltonian terms, one clockwise for $H_\circlearrowright$ and another counterclockwise for $H_\circlearrowleft$.
Only $H_\circlearrowright$ or $H_\circlearrowleft$ \textit{independently} possess the generalized chiral symmetry $\mathscr{C}_3$ in the precise sense of (\ref{eq:genchiralsym}), while $H_0$ does not.
The original $H_0$ is recovered after three consecutive $\Gamma_3$ operations ($H_0=\Gamma_3^3H_0\Gamma^{-3}_3$).
Thus, if we define
\begin{equation}
	H_2:=\Gamma_3H_1\Gamma^{-1}_3=\omega_3^{-2}H_\circlearrowright+ \omega_3^2 H_\circlearrowleft,
\end{equation}
it follows that $H_0=\Gamma_3H_2\Gamma^{-1}_3$ and
\begin{equation}
	H_0+H_1+H_2=0,
	\label{eq:appsumh}
\end{equation}
which formally replicates (\ref{eq:sumcolorsh3}), although $H_0$, $H_1$, and $H_2$ are not different chiral colors of the same Hamiltonian, that is, they do not relate to each other by multiples of $\omega_3$ as in (\ref{eq:hred})-(\ref{eq:hgreen}).
From the cyclic property of the trace of a matrix product, we have
\begin{equation}
	\Tr (H_1)=\Tr (\Gamma_3H_0\Gamma^{-1}_3)=\Tr (H_0),
\end{equation}
and similarly $\Tr (H_2)=\Tr (H_0)$.
From applying the trace to both sides of (\ref{eq:appsumh}) we conclude that $\Tr (H_0)=0$, that is, the eigenvalues of $H_0$ sum to zero.

Let us consider a general $4\times 4$ Hamiltonian of the form
\begin{equation}
	H=
	\begin{pmatrix}
		a&b&c&d
		\\
		e&f&g&h
		\\
		i&j&k&l
		\\
		m&n&o&p
	\end{pmatrix},
	\label{eq:apph4by4}
\end{equation}
where $\{a,b,\dots,p\}\in \mathbb{C}^{16}$.
We can decompose $H$ as
\begin{widetext}
	\begin{eqnarray}
		H&=&H_A+H_B+H_C+H_D+H_E,
		\label{eq:apph4decomp}
		\\
		H_A&=&	
		\begin{pmatrix}
			a&&&
			\\
			&f&&
			\\
			&&k&
			\\
			&&&p
		\end{pmatrix},
		H_B=
		\begin{pmatrix}
			&b&&
			\\
			&&g&
			\\
			&&&l
			\\
			m&&&
		\end{pmatrix},
		H_C=
		\begin{pmatrix}
			&&&d
			\\
			e&&&
			\\
			&j&&
			\\
			&&o&
		\end{pmatrix},
		H_D=
		\begin{pmatrix}
			&&c&
			\\
			&&&h
			\\
			&&&
			\\
			&&&
		\end{pmatrix},
		H_E=
		\begin{pmatrix}
			&&&
			\\
			&&&
			\\
			i&&&
			\\
			&n&&
		\end{pmatrix},
	\end{eqnarray}
\end{widetext}
where all entries not shown are zeros.
The generalized chiral symmetry operator acts in this case as
\begin{equation}
	\Gamma_4H\Gamma^{-1}_4=H_A+\omega_4^{-1}H_B+\omega_4 H_C+\omega_4^{-2}H_D+\omega_4^2H_E,
	\label{eq:appchiralh3}
\end{equation}
which obeys 
\begin{equation}
	\sum\limits_{j=0}^3 \Gamma_4^jH\Gamma^{-j}_4=4H_A,
	\label{eq:appgenchiral}
\end{equation}
with $\Gamma_4^0=\Gamma_4^4=\mathbb{1}_4$.
Note that $\omega_4=e^{i\frac{\pi}{2}}$ is imposed by both $\Gamma_4H_B\Gamma^{-1}_4=\omega_4^{-1}H_B$ and $\Gamma_4H_C\Gamma^{-1}_4=\omega_4H_C$.
Taking the trace on both sides of (\ref{eq:appgenchiral}) leads to
\begin{equation}
	\Tr (\sum\limits_{j=0}^3 \Gamma_4^jH\Gamma^{-j}_4)=4\Tr (H_A)=4\Tr (H),
\end{equation}
which, when $H_A$ is traceless, $\Tr (H_A)=a+f+k+p=0$, implies that the eigenvalues of $H$ sum to zero \cite{Ni2019,Li2020b}. When $H_A=O_{4\times 4}$, where $O_{j\times j}$ is the $j\times j$ zero matrix, then (\ref{eq:appgenchiral}) further reduces to the generalized chiral symmetry proposed in \cite{Li2020b,Li2021}.
The decomposition in (\ref{eq:apph4decomp}) can be straightforwardly generalized to a Hamiltonian of any dimension such that, under the action of $\Gamma_n$, the different components rotate by multiples of $\omega_n$, as illustrated in (\ref{eq:appchiralh3}) for $n=4$.
Therefore $n$ successive applications of $\Gamma_n$ will retrieve the original Hamiltonian, implying that \textit{all} Hamiltonians with a trivial main diagonal obey
\begin{equation}
	\sum_{j=0}^{n-1}\Gamma_n^jH\Gamma^{-j}_n=0.
\end{equation}
However, only a small subset of these Hamiltonians, which includes at least the chiral colored ones with the form of (\ref{eq:hamiltnroot}) (and evidently also their conjugate transposed versions, corresponding to a global inversion of the hopping directions), can be said to enjoy $\mathscr{C}_n$-symmetry in the more stringent sense of (\ref{eq:genchiralsym}) [with $\omega_n^{-1}\to \omega_n$ for the conjugate transpose versions $H(\mathbf{k})\to H^\dagger(\mathbf{k})$], in light of which it can be described as an extension of chiral symmetry to $n$-partite lattices, since it reduces to the usual chiral symmetry for a bipartite ($n=2$) model.

\section{Analogy with Baxter's clock model}
\label{app:baxter}

In order to highlight the parallel than can be drawn between the non-Hermitian $n$-partite models we are considering and the generalized version of the Ising model, we will follow closely below the systematic analysis provided by Fendley \cite{Fendley2014}, to which we refer the reader for further details.
Baxter's clock model \cite{Baxter1989,Baxter1989b,Fendley2012} can be viewed as an extension of the 1D Ising chain where, instead of having ``up'' and ``down'' as the internal spin degree of freedom at each site, the ``spin'' or clock value at each site can take the value $\omega_n^j=e^{i\frac{2\pi}{n}j}$, with $j=0,1,\dots,n-1$.
Its Hamiltonian for an $L$ sites chain reads as
\begin{equation}
	H_{\text{BC}}=\sum\limits_{l=1}^Lt_{2l-1}\tau_l + \sum\limits_{l=1}^{L-1}t_{2l}\sigma_l^\dagger\sigma_{l+1},
	\label{eq:apphbaxter}
\end{equation}
where $\{t_i\}$ is a set of $2L-1$ arbitrary complex coefficients and $\sigma_j=\mathbb{1}_n\otimes\dots\mathbb{1}_n\otimes\sigma\otimes\mathbb{1}_n\dots$ and $\tau_j=\mathbb{1}_n\otimes\dots\mathbb{1}_n\otimes\tau\otimes\mathbb{1}_n\dots$ are operators acting at site $l=1,2,\dots,L$ \cite{Albertini1989} through the local operators \cite{Mittag1971}
\begin{equation}
	\sigma=\begin{pmatrix}
		1
		\\
		&\omega_n
		\\
		&&\omega_n^2
		\\
		&&&\ddots
		\\
		&&&&\omega_n^{n-1}
	\end{pmatrix},
	\tau=\begin{pmatrix}
		&&&&1
		\\
		1
		\\
		&1
		\\
		&&\ddots
		\\
		&&&1
	\end{pmatrix},
	\label{eq:appsigmatau}
\end{equation}
where all entries not shown are zeros and the local basis at site $l$ spans $\{\ket{\omega_n^{j}}\}$, with $j=0,1,\dots,n-1$.
Clearly, $\sigma$ measures the clock value, $\sigma\ket{\omega_n^{j}}=\omega_n^{j}\ket{\omega_n^{j}}$, and $\tau$ is the shifting operator acting as $\tau\ket{\omega_n^{j}}=\ket{\omega_n^{j+1}}$.
The Ising model is recovered for $n=2$, where $\sigma$ and $\tau$ reduce to the $\sigma_z$ and $\sigma_x$ Pauli matrices, respectively. 
For $n>2$, the one-site term of $H_{\text{BC}}$ in (\ref{eq:apphbaxter}) generalizes the spin flipping term, while the two-site term represents a generalized nearest-neighbor interaction.
Baxter's model also enjoys a generalized chiral symmetry defined as
\begin{eqnarray}
	\mathscr{C}_n:\ \ \ \Sigma_n H_{\text{BC}}\Sigma^{-1}_n&=&\omega_nH_{\text{BC}},
	\label{eq:appgenchiralsym}
	\\
	\Sigma_n&=&\prod\limits_{j=1}^L\sigma_j\prod\limits_{i=1}^L\tau_i^{-i},
\end{eqnarray} 
with $\Sigma_n$ a unitary operator whose action cycles the chiral colors of $H_{\text{BC}}$ \footnote{This symmetry was labeled as a generalized charge conjugation symmetry (also known as particle-hole symmetry) in \cite{Fendley2014}. However, charge conjugation is defined through an antiunitary operator, whereas $\Sigma_n$ is unitary and therefore is rather the operator describing a generalized chiral symmetry.}.
The presence of $\mathscr{C}_n$-symmetry similarly imposes that the spectrum of $H_{\text{BC}}$ be formed by sequences of the form $\{E,\omega_nE,\omega_n^2E,\dots,\omega_n^{n-1}E\}$, each summing to zero [see discussion below (\ref{eq:schrodgenchiral})].
Notice that (\ref{eq:appgenchiralsym}) is equivalent to (\ref{eq:genchiralsym}) if the direction of all hopping terms is switched, such that $H(\mathbf{k})\to H^\dagger(\mathbf{k})$ in (\ref{eq:hamiltnroot}).

An even more direct analogy can be made with the ``$\omega$-commutation'' relation \cite{Baxter1989} between the local $\sigma$ and $\tau$ operators in (\ref{eq:appsigmatau}), which we write here in a slightly different fashion for comparison purposes,
\begin{equation}
	\sigma\tau^{\mathsmaller T}\sigma^{-1}=\omega_n^{-1}\tau^{\mathsmaller T},
	\label{eq:appwcommut}
\end{equation}
which can be rewritten as a phase commutator of the form of (\ref{eq:phasecomm}) as $[\sigma,\tau^{\mathsmaller T}]_{\phi_n}=0$.
Notice that $\tau^{\mathsmaller T}$ has the same general form of $H(\mathbf{k})$ in (\ref{eq:hamiltnroot}), where the ones are replaced by the $h_j$ rectangular matrices of different sizes. Similarly, $\sigma$ closely resembles $\Gamma_n$ in (\ref{eq:genchiraloperator}), where each diagonal power of $\omega_n^l$, with $l=0,1,\dots,n-1$, is enlarged into a diagonal square block of dimension $d_{l+1}$, that is, $\omega_n^l\to\omega_n^l\mathbb{1}_{d_{l+1}}$.
In the interest of keeping up with this analogy, the bulk Hamiltonian of the non-Hermitian $n$-partite models with the form of (\ref{eq:hamiltnroot}) can be viewed as a generalized version of the shift operator of Baxter's clock model, where each SL corresponds to a different clock value, and the dimension $d_j$ of SL$_j$, \textit{i.e.}, the number of sites at SL$_j$, can be viewed as counting the internal degrees of freedom of each clock value, connected between adjacent SLs in a unidirectional cyclic fashion.

It should also be stressed that, when mapping $\tau^{\mathsmaller T}\to H(\mathbf{k})$, a new ingredient is added to the system that is at the heart of the generalized Lieb's theorem of (\ref{eq:genlieb}).
Namely, the finite scalar entries of $\tau^{\mathsmaller T}$ are converted into rectangular matrices $h_j$ with different dimensionalities in general, such that diagonal square blocks $H_j$ of different sizes $d_j$ are obtained for $H^n(\mathbf{k})$, whereas one trivially gets $(\tau^{\mathsmaller T})^n=\mathbb{1}_n$.
From (\ref{eq:genlieb}), a nontrivial number of zero-energy FBs is precisely the combined result of finite sublattice imbalances (first term) and/or the existance of zero-energy FBs in the $H_1$ block (second term), both of which vanish for $(\tau^{\mathsmaller T})^n$.
At this point the analogy stops, since the generalized Lieb's theorem cannot be revealed by the mathematical structure of Baxter's clock model, but can be derived from the properties of the non-Hermitian $n$-partite models we considered.

\section{Comments on the Lieb's theorem}
\label{app:4rootcl}

Here, we show that Lieb's theorem, which states that the number of zero-energy FBs in a bipartite system is given by the sublattice imbalance, while correct for Hermitian systems, fails to account for the extra zero-energy FBs that appear in certain non-Hermitian lattices.
We show below an example of such a model, further illustrating the validity of the generalized Lieb's theorem expressed in (\ref{eq:genlieb}) already at the bipartite ($n=2$) level.
\begin{figure}[ht]
	\begin{centering}
		\includegraphics[width=0.48 \textwidth,height=3.cm]{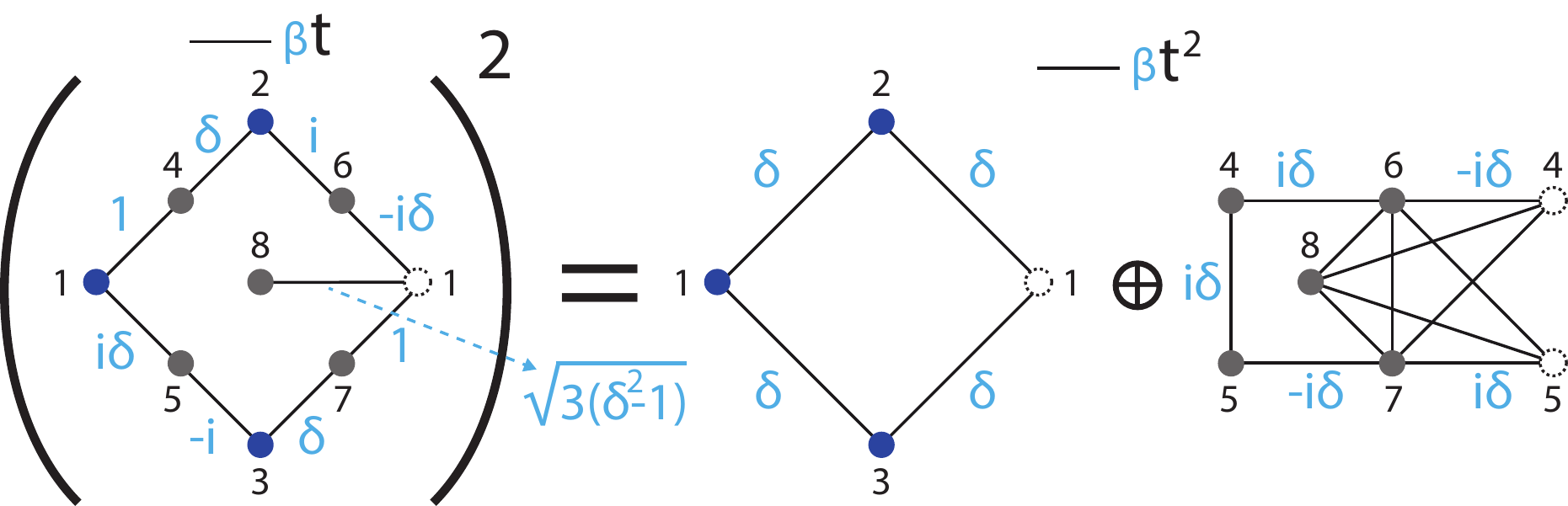} 
		\par\end{centering}
	\caption{(Left) Unit cell of the non-Hermitian four-root topological insulator with different prefactors $\beta$ at different hopping terms. The non-Hermiticity comes from the Peierls phases picked up by the hopping terms, which are the same in both directions. (Right) When squared, the model at the left leads to two decoupled models, namely a diamond chain in the blue SL$_1$ and a two-leg ladder in the gray SL$_2$ (only some hopping terms are indicated for the latter). Open sites belong to the respective adjacent unit cells.}
	\label{fig:4rootti}
\end{figure}

\begin{figure*}[t]
	\begin{centering}
		\includegraphics[width=0.975 \textwidth,height=4.1cm]{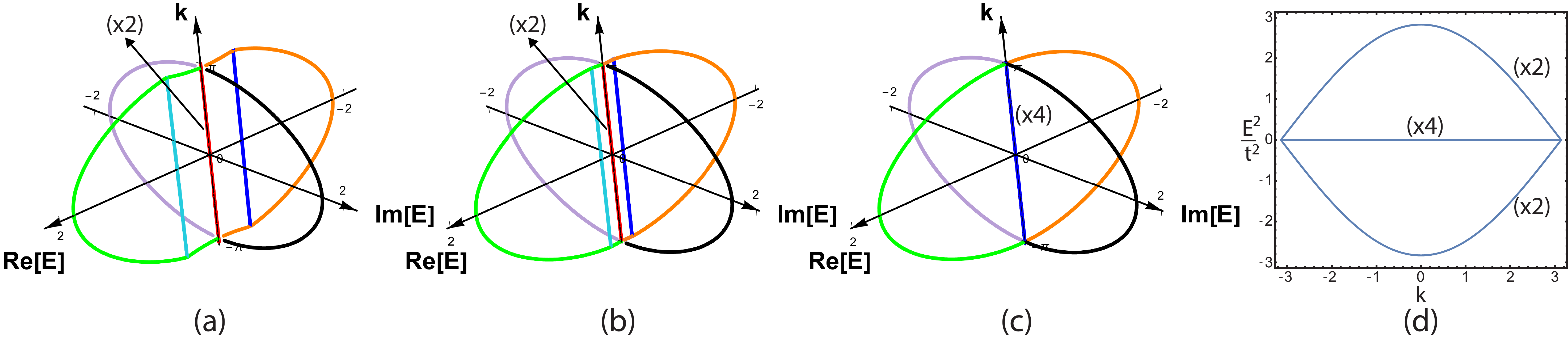} 
		\par\end{centering}
	\caption{Complex energy spectrum as a function of the momentum obtained from diagonalizing the Hamiltonian defined in (\ref{eq:apph4root}) for (a) $\delta=1.1$, (b) $\delta=1.01$, and (c) $\delta=1$. (d) Squared energy spectrum as a function of the momentum obtained by squaring the Hamiltonian with the parameters of (c), which is purely real. In all plots, ($\times j$) indicates the $j$-fold degeneracy of the respective band.}
	\label{fig:espectrumdc2}
\end{figure*}
The model considered here, with the unit cell depicted in Fig.~\ref{fig:4rootti} at the left, can be viewed as a non-Hermitian variation on the 1D four-root topological insulator studied in \cite{Marques2021}.
The bulk Hamiltonian, parametrized by the real $\delta$ factor included at some hopping parameters, reads as
\begin{eqnarray}
	H(k,\delta)&=&t\begin{pmatrix}
		O_{3\times 3}&h(k,\delta)
		\\
		h^T(-k,\delta)&O_{5\times 5}
	\end{pmatrix},
	\label{eq:apph4root}
	\\
	h(k,\delta)&=&
	\begin{pmatrix}
		1&i\delta&-i\delta e^{-ik}&e^{-ik}&\sqrt{3(\delta^2-1)}e^{-ik}
		\\
		\delta&0&i&0&0
		\\
		0&-i&0&\delta&0
	\end{pmatrix},
\end{eqnarray}
The non-Hermiticity comes from the finite Peierls phases at some of the hopping terms, which are the same in both directions.
Squaring $H(k,\delta)$ in (\ref{eq:apph4root}) leads to
\begin{eqnarray}
	H^2(k,\delta)&=&t^2\begin{pmatrix}
		H_1(k,\delta)&O_{3\times 5}
		\\
		O_{5\times 3}&H_2(k,\delta)
	\end{pmatrix},
	\label{eq:apph4root2}
	\\
	H_1(k,\delta)&=&(\delta^2-1)\mathbb{1}_3+\delta
	\begin{pmatrix}
		0&1+e^{-ik}&1+e^{-ik}
		\\
		1+e^{ik}&0&0
		\\
		1+e^{ik}&0&0
	\end{pmatrix},
	\label{eq:apphdiamond}
\end{eqnarray}
where $H_1(k,\delta)$ models the diamond chain with the unit cell depicted at the blue SL$_1$ at the right-hand side of Fig.~\ref{fig:4rootti}, which is known to host a flat band with the energy of its diagonal term \cite{Pelegri2019,Pelegri2019b,Kremer2020,Pelegri2020}, that is, $E_{\text{FB}}=\delta^2-1$. 
The full expression of the pseudo-Hermitian block $H_2(k,\delta)$, modeling a chain of the form of the gray SL$_2$ at the right-hand side of Fig.~\ref{fig:4rootti}, is omitted here for simplicity. 
In the language of \cite{Marques2021}, it corresponds to a topologically featureless residual block with shared spectral properties with the relevant $H_1(k,\delta)$ block.

In Figs.~\ref{fig:espectrumdc2}(a)-\ref{fig:espectrumdc2}(c), we plot the complex energy spectrum of $H(k,\delta)$ in (\ref{eq:apph4root}) for three decreasing values of $\delta$.
For all three cases, there are two zero-energy FBs originating from sublattice imbalance, in accordance with Lieb's theorem.
However, two extra FBs with symmetric energies, directly obtained by taking the square-root of the diagonal term in (\ref{eq:apphdiamond}), \textit{i.e.}, $E_{\text{FB},\pm}=\pm\sqrt{\delta^2-1}$, are present in the spectra and can be seen to coalesce with the other two FBs as $\delta\to 1$ (they evolve in the imaginary energy axis for $|\delta|<1$).
Therefore, in Fig.~\ref{fig:espectrumdc2}(c) the system displays two extra zero-energy FBs not accounted for by Lieb's theorem.
Their appearance comes from the fact that, for the squared Hamiltonian $H^2(k,1)$, whose energy spectrum is shown in Fig.~\ref{fig:espectrumdc2}(d), the smaller $H_1(k,1)$ block \textit{itself} has a zero-energy FB ($\#_{\text{FB}}^{H_1}=1$ due to the sublattice imbalance within SL$_1$), since its diagonal term vanishes at $\delta=1$, which must be shared by the $H_2(k,1)$ block also due to the isospectral properties (up to the zero-energy FBs already accounted for by the sublattice imbalance) between the diagonal blocks, as discussed in the main text. 
For $\delta=1$, the total number of zero-energy FBs obtained from the generalized Lieb's theorem is
\begin{eqnarray}
	\#_{\text{FB}}=d_2-d_1+2\#_{\text{FB}}^{H_1}=5-3+2=4.
	\label{eq:appcgenlieb4root}
\end{eqnarray}

There is a simple reason why the term proportional to $\#_{\text{FB}}^{H_1}$ is absent from Lieb's theorem.
It relates to the fact that it applies to Hermitian systems, where no zero-energy FBs can be present in $H_1$, apart from the trivial case where decoupled sites are present within the unit cell, which can allways be chosen to belong to the larger sublattice (notice, e.g., that site 8 becomes decoupled for the left model of Fig.~\ref{fig:4rootti} when $\delta=1$, at which point one can ascribe it to either sublattice).
When finite Hermitian couplings between sites in SL$_1$ and SL$_2$ are considered for a bipartite system, the diagonal terms of the squared Hamiltonian are necessarily positive \cite{Ezawa2020,Marques2021,Marques2021b}.
If $H_1$ is itself bipartite, then it has $\#_{\text{FB}}^{H_1}>0$ coming from its sublattice imbalance, which are replicated in $H_2$, but with a finite energy given by its diagonal term $c$ [with, e.g., $c=\delta^2-1$ for the $H_1$ block in (\ref{eq:apphdiamond})].
In the original model, this translates in the appearance of $\#_{\text{FB}}^{H_1}$-fold degenerate FBs at $E=\pm\sqrt{c}$. 
These FBs are pushed in pairs to zero energy as $c\to 0$.
The only way this can be achieved is by adding \textit{negative} contributions to the diagonal term $c$ of the squared model which, in turn, requires the inclusion of non-Hermitian hopping terms in the original model, as we exemplified in (\ref{eq:apph4root}) by considering non-Hermitian Peierls phases at some couplings.
This demonstrates, in short, that Lieb's theorem needs to be generalized, not only for the $n$-partite systems considered in the main text, with $n>2$, but also already for non-Hermitian bipartite systems, where extra zero-energy FBs originate from those that may be present in the $H_1$ squared block.

If one considers the diamond chain depicted in the middle of Fig.~\ref{fig:4rootti}, but introducing now a $\pi$-flux per plaquette, such that the dispersive bands above and below the zero-energy FB [see Fig.~\ref{fig:espectrumdc2}(d)] also become FBs with symmetric finite energies \cite{Pelegri2019,Pelegri2019b,Pelegri2020}, then the same procedure followed for the four-root model of Fig.~\ref{fig:4rootti} can be applied.
Namely, by introducing carefully selected non-Hermitian phases at some of the couplings, while keeping the $\pi$-flux pattern per plaquette \cite{Leykam2017,Zhang2020}, the symmetric finite energy bands can be pushed to zero-energy, which corresponds, in its squared Hamiltonian, to lowering the energy of the smallest block (a single FB) to zero.
In this scenario, one obtains a completely trivial energy landscape made of three zero-energy FBs, as was recently shown by Ding \textit{et al.} \cite{Ding2021} in a non-Hermitian system of coupled resonators.
As for the case of the four-root model analyzed in this appendix that led to (\ref{eq:appcgenlieb4root}), our results also provide a full account of  the two extra FBs appearing in this diamond chain system.

\section{Comments on defective Hamitonians}
\label{app:defective}

The minimum number of LIEs within the set of zero-energy FBs follows from a simple argument. The adjacency graph of $H(\mathbf{k})$ is a directed graph with only outgoing links from SL$_{j}$ to SL$_{j-1}$, where $j=1, \cdots, n$ and $j=0\to j=n$ from the periodicity. Assuming that the number of dispersive bands of $H^n(\mathbf{k})$ is given by the dimension of the smallest block $H_1$, that is , it is $nd_1$ [where $n$ reflects the $n$-fold degeneracy of $H^n(\mathbf{k})$], then the incoming hopping terms to each site $i_1$ in SL$_{1}$ determine a particular state $\ket{\psi_{i_1}^{(2)}}$ in SL$_{2}$. We can construct a basis for SL$_{2}$ applying a Gram-Schmidt orthonormalization to the states $\{\ket{\psi_{i_1}^{(2)}}\}$ (note this set spans a subspace of SL$_{2}$  of dimension $d_1$) and choosing an arbitrary set of orthonormal basis states (between themselves and to the set  $\{\ket{\psi_{i_1}^{(2)}}\}$) that completes the basis of SL$_{2}$. If we draw the adjacency graph of $H(\mathbf{k})$ in this basis, the latter set of nodes will have no outgoing links and that will generate $d_2-d_1$ zero-energy FBs with LIEs.
\begin{figure}[ht]
	\begin{centering}
		\includegraphics[width=0.34 \textwidth,height=3.5cm]{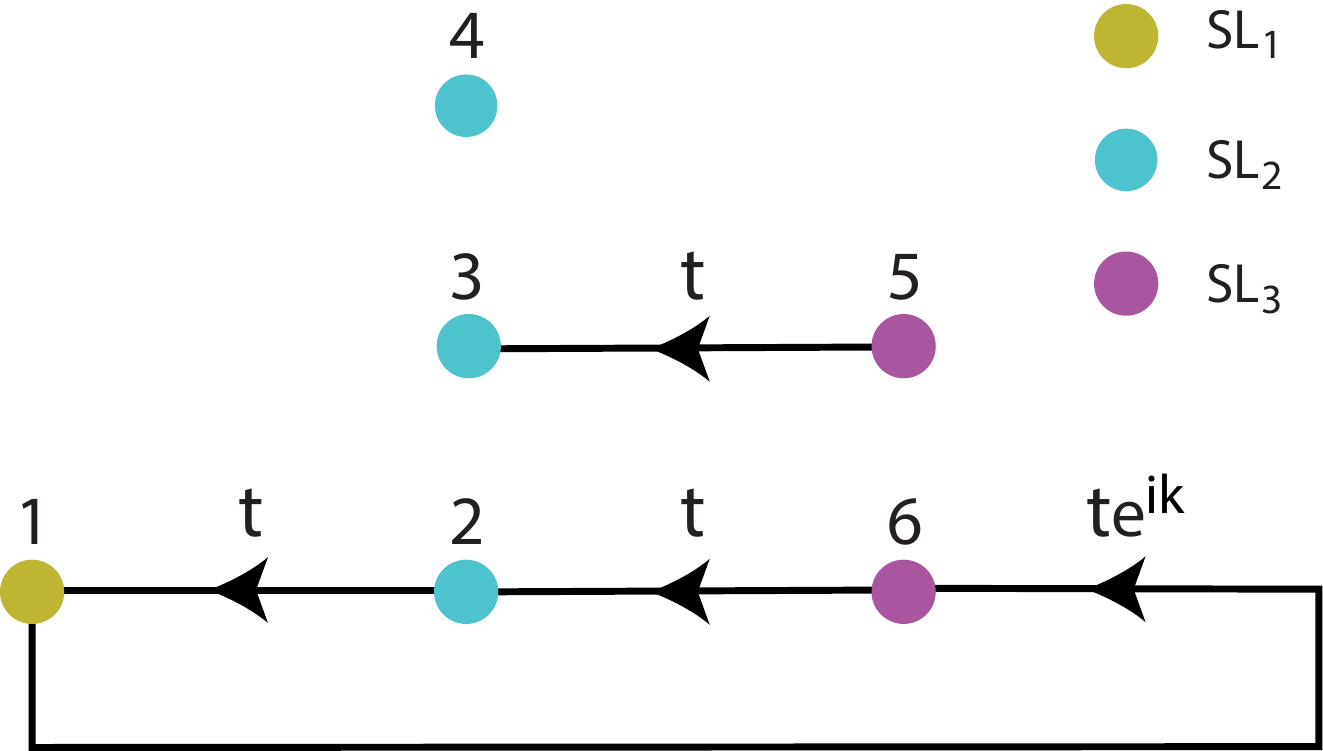} 
		\par\end{centering}
	\caption{Unit cell of a tripartite defective Hamiltonian. The hopping terms are unidirectional, acting only in the direction of the arrows.}
	\label{fig:1a}
\end{figure}

This argument can be extended to any  Hamiltonian block between a pair of consecutive sublattices. Three situations can occur: (i) $d_j > d_{j-1}$, (ii) $d_{j} = d_{j-1}$, and (iii) $d_{j} < d_{j-1}$. For the two latter cases, no nodes of SL$_{j}$ without outgoing links can be obtained with the procedure described above. When $d_{j} > d_{j-1}$, then the same reasoning will generate  $d_{j}-d_{j-1}$ nodes in SL$_{j}$ without outgoing links. So the minimum number of LIEs in the set of zero-energy FBs is 
\begin{equation}
	\#_{\text{LIEs}}^{\text{min}}=\sum\limits_{j=2}^n \text{Max}(d_j-d_{j-1},0).
	\label{eq:genlieb2}
\end{equation}

Basically, the argument above states that is possible to rotate the basis within each sublattice in such a way that the number of sites in the shortest section (smallest sublattice) in closed loops of the adjacency graph gives the number of dispersive bands and the number of endpoints of open paths gives the number of LIEs of the set of zero-energy FBs, while the total number of FBs in this set is always given by the generalized Lieb's theorem in (\ref{eq:genlieb}).
In the example of Fig.~\ref{fig:1a}, while the model, through (\ref{eq:genlieb}), has $\#_{\text{FB}}=3$ zero-energy FBs, this set only counts
\begin{equation}
	\#_{\text{LIEs}}^{\text{min}}=\text{Max}(d_2-d_{1},0)+\text{Max}(d_3-d_{2},0)=2+0=2.
\end{equation} 
The defectiveness of the model comes in this case from the decoupled cluster within the unit cell involving sites 3 and 5, which yields two FBs but only one LIE, while the loop accounts for the three dispersive bands and decoupled site 4 for the other zero-energy FB.
This defectiveness can be viewed as the skin effect that takes place for the decoupled cluster within each unit cell, which together form a set of decoupled nonreciprocal dimers in real-space.

\bibliography{genlieb}

\end{document}